\documentclass[aip,jap,preprint]{revtex4-1}

\usepackage{graphicx}
\usepackage{color}
\usepackage{multirow}
\usepackage{dcolumn}
\usepackage{subfigure}
\usepackage{textcomp}

\usepackage{lineno}

\begin{document}


\title{Low-Temperature Scintillation Properties of CaWO$_4$ Crystals for Rare-Event Searches}



\author{M.~v. Sivers}
\email{moritz.vonsivers@lhep.unibe.ch}
\affiliation{Physik-Department, Technische Universit\"{a}t M\"{u}nchen, 85748 Garching, Germany}
\affiliation{present address: Albert Einstein Center for Fundamental Physics, University of Bern, 3012 Bern, Switzerland}
\author{M.~Clark}
\author{P.C.F.~Di Stefano}
\affiliation{Department of Physics, Engineering Physics \& Astronomy, Queen's University, Kingston, Ontario, Canada, K7L 3N6}
\author{A.~Erb} 
\affiliation{Physik-Department, Technische Universit\"{a}t M\"{u}nchen, 85748 Garching, Germany}
\affiliation{Walther-Mei{\ss}ner-Institute for Low Temperature Research, 85748 Garching, Germany}
\author{A.~G\"{u}tlein} 
\affiliation{Physik-Department, Technische Universit\"{a}t M\"{u}nchen, 85748 Garching, Germany}
\affiliation{present address: Institut f\"ur Hochenergiephysik der \"Osterreichischen Akademie der Wissenschaften, A-1050 Wien, Austria \\
and Atominstitut, Vienna University of Technology, A-1020 Wien, Austria}
\author{J.-C.~Lanfranchi}
\author{A.~M\"unster}
\affiliation{Physik-Department, Technische Universit\"{a}t M\"{u}nchen, 85748 Garching, Germany}
\author{P.~Nadeau}
\author{M.~Piquemal}
\affiliation{Department of Physics, Engineering Physics \& Astronomy, Queen's University, Kingston, Ontario, Canada, K7L 3N6}
\author{W.~Potzel} 
\affiliation{Physik-Department, Technische Universit\"{a}t M\"{u}nchen, 85748 Garching, Germany}
\author{S.~Roth} 
\affiliation{Physik-Department, Technische Universit\"{a}t M\"{u}nchen, 85748 Garching, Germany}
\affiliation{present address: Department of Physics, Engineering Physics \& Astronomy, Queen's University, Kingston, Ontario, Canada, K7L 3N6}
\author{K.~Schreiner}
\affiliation{Department of Physics, Engineering Physics \& Astronomy, Queen's University, Kingston, Ontario, Canada, K7L 3N6}
\author{R.~Strauss} 
\affiliation{Physik-Department, Technische Universit\"{a}t M\"{u}nchen, 85748 Garching, Germany}
\affiliation{present address: Max-Planck-Institut f\"ur Physik, D-80805 M\"unchen, Germany}
\author{S.~Wawoczny} 
\author{M.~Willers}
\author{A.~Z\"{o}ller} 
\affiliation{Physik-Department, Technische Universit\"{a}t M\"{u}nchen, 85748 Garching, Germany}


\date{\today}

\begin{abstract}
In prospect of its application in cryogenic rare-event searches, we have investigated the low-temperature scintillation properties of CaWO$_4$ crystals down to 3.4\,K under $\alpha$ and $\gamma$ excitation.
Concerning the scintillation decay times, we observe a long component in the ms range which significantly contributes to the light yield below 40\,K.
For the first time we have measured the temperature dependence of the $\alpha$/$\gamma$-ratio of the light yield. This parameter which can be used to discriminate $\alpha$ and $\gamma$ events in scintillating bolometers is found to be $\sim$8-15\% smaller at low temperatures compared to room temperature.
\end{abstract}


\maketitle

\section{Introduction}
Inorganic scintillators are widely used as detectors in high-energy physics and medical imaging. The operation of scintillators at low temperatures has the advantage that the light yield of most materials with intrinsic luminescence increases with decreasing temperature \cite{mikhailik10}. In addition, scintillating crystals at mK temperatures can be operated as bolometers measuring the phonon signal produced by a particle interaction. By the simultaneous detection of the generated scintillation light an efficient discrimination between different particles can be achieved \cite{bobin97,alessandrello98,meunier99}. Such scintillating bolometers are being used in rare-event searches like dark matter direct detection experiments or searches for the neutrinoless double beta decay \cite{angloher12, cardani12, beeman12, bhang12}. \\
CaWO$_4$ is an interesting material for scintillating bolometers. Together with ZnWO$_4$ and CdWO$_4$ it shows the highest light yield at low temperatures among all tungstates and molybdates \cite{mikhailik10}. In addition, the high Debye temperature \cite{senyshyn04} leads to a good energy resolution and low energy threshold for the phonon signal \cite{angloher14a,angloher15a}. Finally, it has been shown that by careful material selection crystals with high radiopurity can be grown \cite{munster14}.
Scintillating CaWO$_4$ crystals are currently being used as bolometers in the direct dark matter search CRESST-II (Cryogenic Rare Event Search with Superconducting Thermometers) \cite{angloher12}. It is planned that similar detectors will be operated in the future EURECA (European Rare Event Calorimeter Array) project \cite{angloher14}. This experiment will comprise a multi-material target of up to 1\,ton consisting of Ge and CaWO$_4$. Furthermore, CaWO$_4$ has also been proposed as a possible candidate for the search of the neutrinoless double beta decay of $^{48}$Ca \cite{zdesenko05a}.\\ 
Although it is one of the oldest known scintillators, there are only few studies of the scintillation properties of CaWO$_4$ below the temperature of liquid nitrogen (77\,K). The evolution of the light yield, defined as the number of photons generated per unit energy, and scintillation decay times down to 10\,K were first measured in Ref.~\onlinecite{beard62}. A more recent investigation of the scintillation characteristics down to mK temperatures is given in Ref.~\onlinecite{mikhailik07}. In Ref.~\onlinecite{mikhailik07} the decay times for $\alpha$ particles and $\gamma$ quanta as well as the light yield, albeit only for $\alpha$ particles, were determined as a function of temperature. Up to now, Ref.~\onlinecite{mikhailik07} has been the only systematic study of the scintillation properties of CaWO$_4$ below 10\,K. Furthermore, there is yet no measurement of the temperature dependence of the light yield under $\gamma$-ray excitation.\\
In the present paper we have determined the scintillation decay times as well as the light yield for both $\alpha$ particles and $\gamma$ quanta between 320\,K and 3.4\,K for two CaWO$_4$ crystals from different suppliers. A direct comparison of the light yield from $\alpha$ and $\gamma$ events allows us for the first time to determine the temperature dependence of the $\alpha$/$\gamma$-ratio of the light yield. This ratio allows discrimination between $\alpha$ and $\gamma$ events with scintillating bolometers and is therefore important to reduce the background in experiments searching for dark matter or the neutrinoless double beta decay \cite{bobin97,alessandrello98}.

\section{Experimental setup}
\begin{figure}
\includegraphics[width=8.5cm]{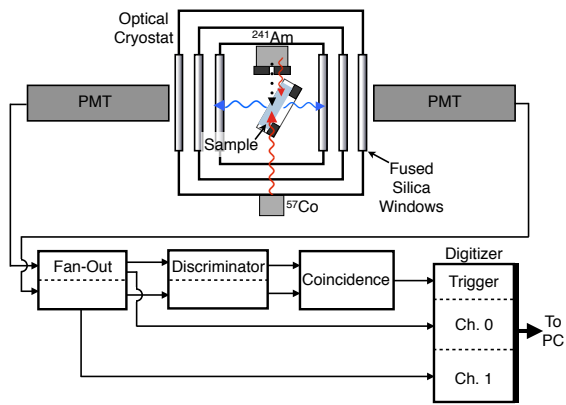}
\caption{\label{fig:cryopt_setup} Schematic picture of the setup used for the measurements. The crystal sample is mounted in an optical cryostat and irradiated by a built-in $^{241}$Am source as well as an optional external $^{57}$Co source. The scintillation light is detected by two PMTs mounted outside of the cryostat. The signals of both PMTs are digitized by an 8-bit ADC with a sampling frequency of 1\,GHz. The trigger requires a coincident signal in both PMTs.}
\end{figure}

The setup used for the measurements is shown in Fig.~\ref{fig:cryopt_setup} and is described in detail in Ref.~\onlinecite{verdier12}. It is based on an optical cryostat cooled by a Gifford-McMahon cryocooler and can reach a base temperature of 3.2 K in about 3.5 hours. The main advantage of this setup is the compact geometry where about 20\% of the solid angle is covered by two photomultiplier tubes (PMTs) mounted outside the cryostat, leading to an improved light collection compared to most optical cryostats. 
A collimated $^{241}$Am source, mounted inside the cryostat, emits $\alpha$ particles as well as 60\,keV $\gamma$-rays. In some measurements we also recorded spectra using an external $^{57}$Co $\gamma$-ray source (122\,keV).
The scintillation light is detected by two 1" Hamamatsu R7056 PMTs with a maximum quantum efficiency QE$\approx$25\% at 420\,nm. This is well matched to the scintillation spectrum of CaWO$_4$ peaking at about the same wavelength \cite{mikhailik10}.
The PMT pulses are recorded by a National Instruments PXI 5154 8-bit digitizer with a sampling rate of 1\,GHz. In an online analysis based on Labview only the samples that exceed a certain threshold above the baseline are saved to disk. The acquisition window was varied between 200\,$\mu$s at room temperature and 3\,ms at the lowest temperature (3.4\,K) according to the expected change of the decay times \cite{mikhailik10}. Between 5\% and 15\% of this acquisition window was used as a pre-trigger. The trigger was provided by a hardware coincidence logic which requires that both PMTs trigger within a coincidence window of 20\,$\mu$s.
The measurement is based on the Multiple Photon Counting Coincidence (MPCC) technique which allows the simultaneous measurement of light yield and decay time \cite{kraus05}.
In order to reject spurious events, such as pile-up and pre-trigger events, some data selection cuts are applied \cite{kraus05,verdier12}.
To obtain the energy spectrum, the integral of each scintillation pulse is calculated while the pulse shape is built using the arrival time and amplitude of every recorded sample in an event. \\
The samples were two CaWO$_4$ crystals with dimensions 5$\times$5$\times$1\,mm$^3$, where the 5$\times$5\,mm$^2$ surface facing the $^{241}$Am source was optically polished and the opposite surface was mechanically roughened to reduce light trapping.
Both of the crystals were produced by the Czochralski method. The first sample was grown in the crystal laboratory of the Technische Universit\"at M\"unchen (TUM) \cite{erb13} while the other one was provided by the General Physics Institute (GPI) of the Russian Academy of Science in Moscow. 

\section{Results and discussion}
\subsection{Average Pulse Shapes}
\label{sec:pulse_shapes}
\begin{figure*}
\subfigure[]{\includegraphics[width=8.cm]{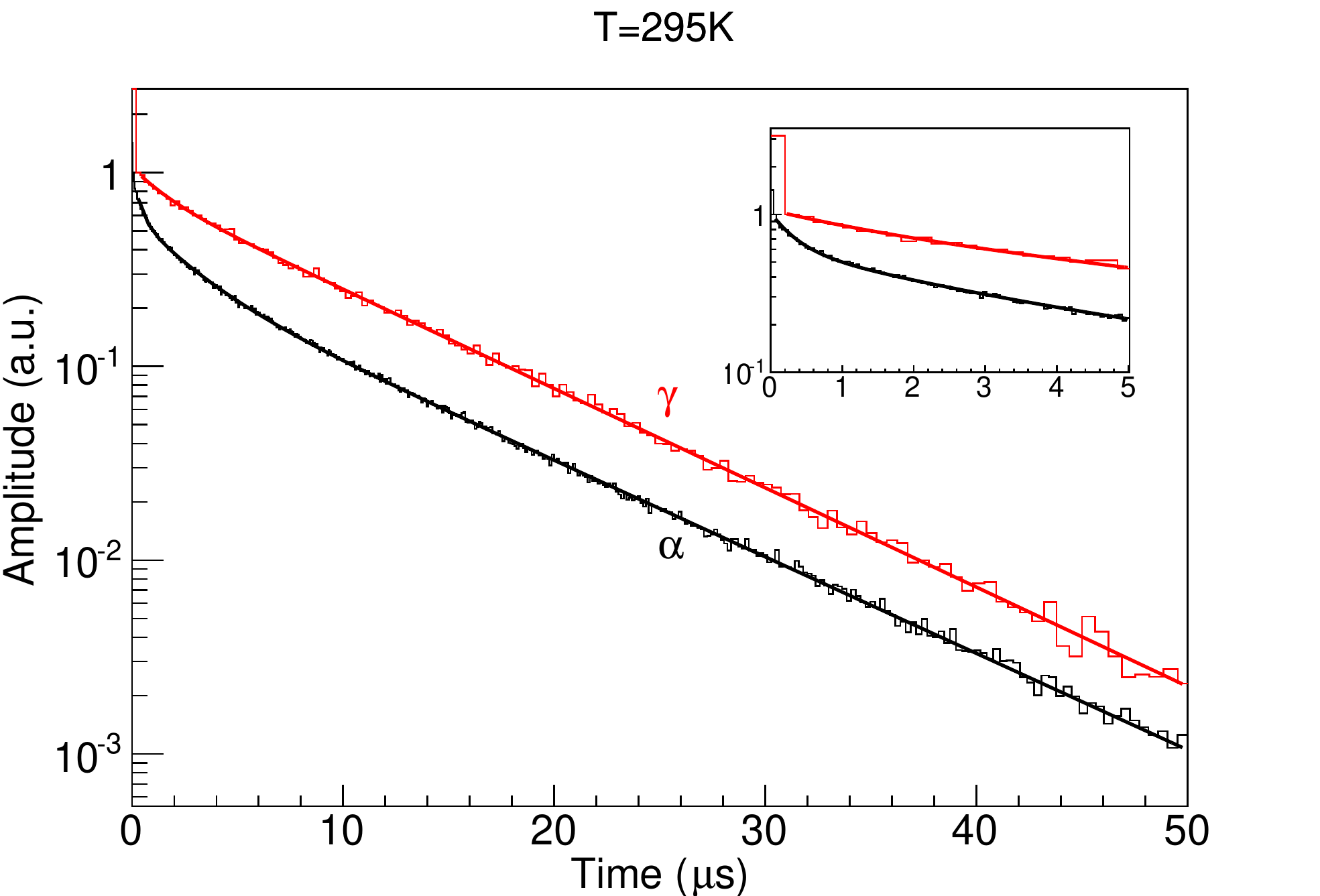}}
\subfigure[]{\includegraphics[width=8.cm]{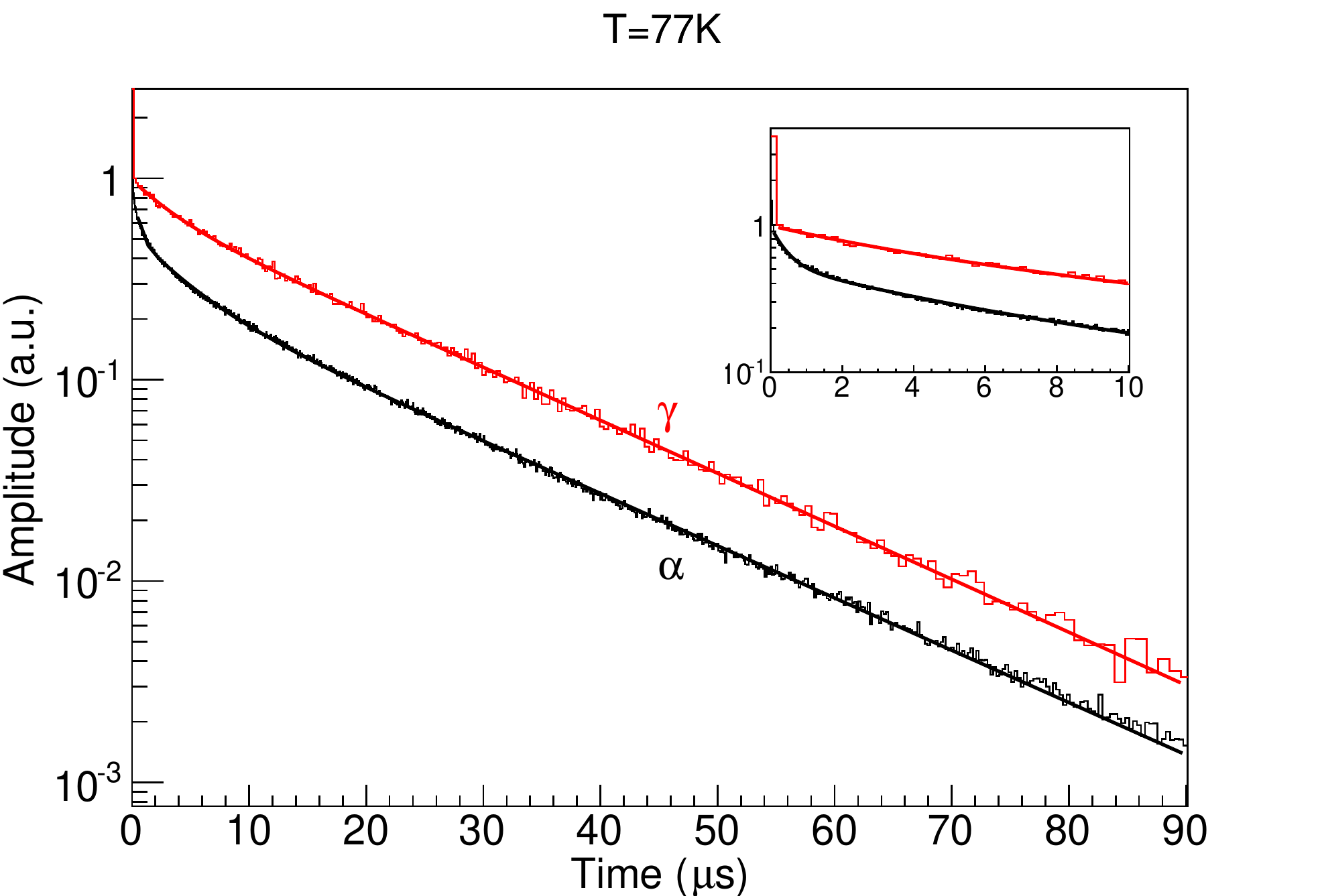}}
\subfigure[]{\includegraphics[width=8.cm]{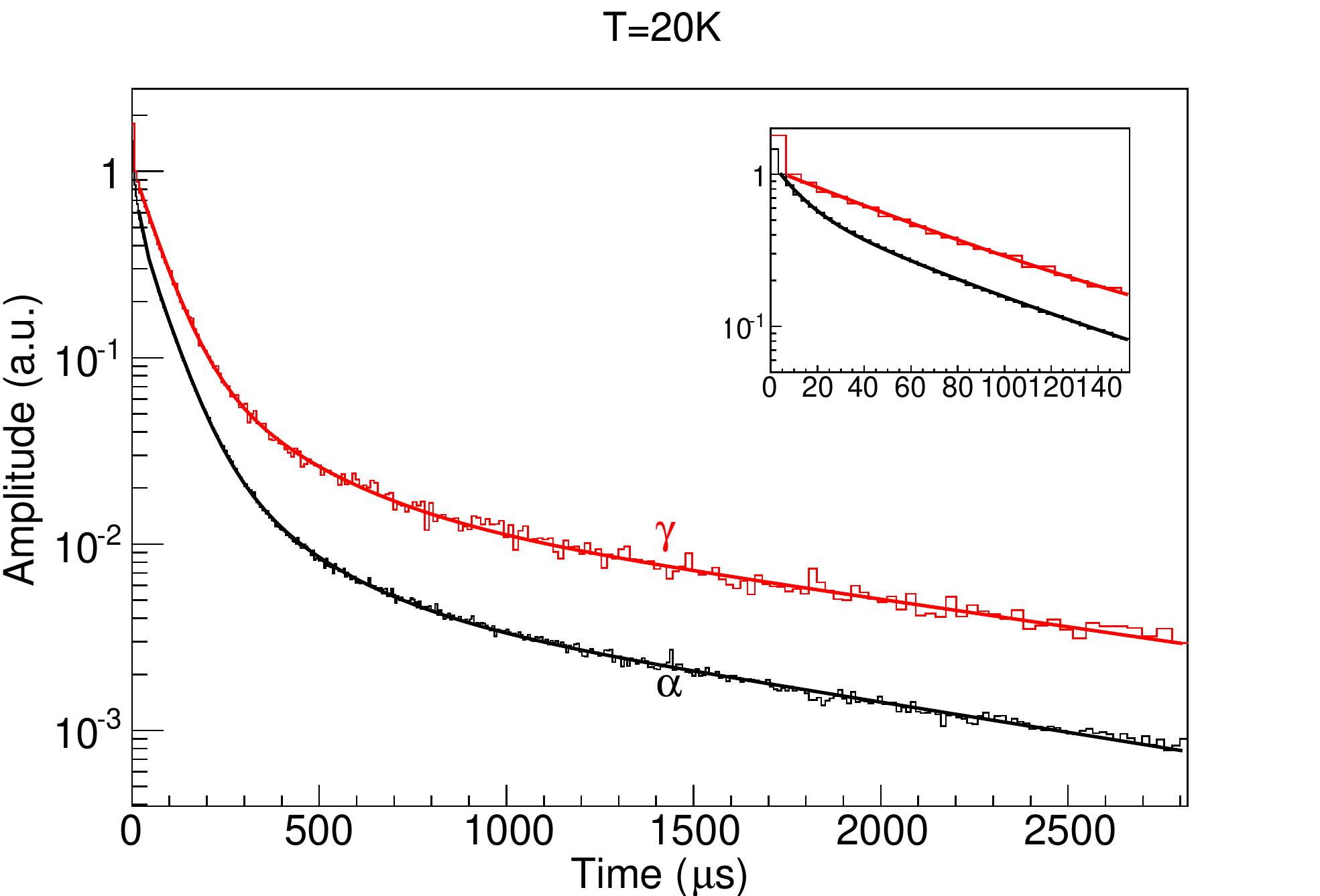}}
\subfigure[]{\includegraphics[width=8.cm]{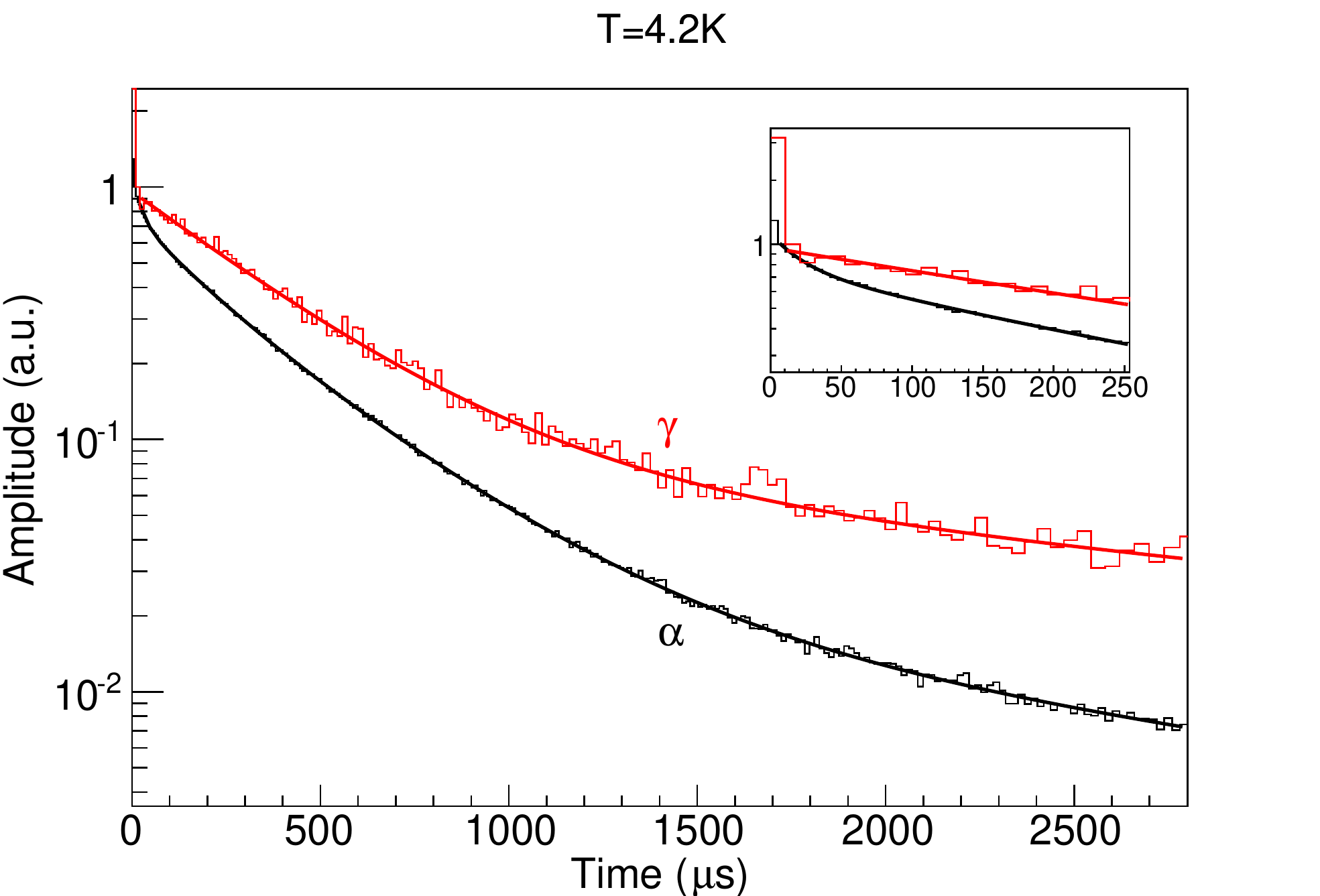}}
\caption{\label{fig:pulses_TUM} Average pulse shapes for $^{241}$Am $\alpha$ and $\gamma$ events at temperatures of 295\,K (a), 77\,K (b), 20\,K (c) and 4.2\,K (d) measured with the TUM crystal (note variable bin size). The amplitude was normalized to the second bin. The inset shows a zoom into the region of the beginning of the pulse. All pulses were fitted with a sum of exponentials (see Eq.~(\ref{eq:pulses})). The fit values are shown in Tab.~\ref{tab:decaytimes}.}
\end{figure*}

Fig.~\ref{fig:pulses_TUM}(a)-(d) shows the pulse shapes at different temperatures for 4.7\,MeV $\alpha$ and 60\,keV $\gamma$ events recorded with the TUM crystal. The pulse shapes were averaged over all scintillation events from the full-energy peaks. It has to be pointed out that the first photon was always taken as the start of a scintillation event since the real start time is unknown. This means that the first bin in the average pulse shape gets overestimated. The effect is more prominent for scintillation events with a low number of photons and can be seen in the average pulses of the 60\,keV $\gamma$-rays shown in Fig.~\ref{fig:pulses_TUM}(a)-(d) (inset). To obtain the scintillation decay times $\tau_i$ each pulse was fitted with a sum of exponentials according to Eq.~(\ref{eq:pulses}) while the first bin was excluded from the fit.
\begin{equation}
f(t)=\sum_i{\frac{n_i}{\tau_i}\exp(-t/\tau_i)}
\label{eq:pulses}
\end{equation}
In Eq.~(\ref{eq:pulses}), $n_i$ stands for the number of photons in the scintillation decay component with decay time $\tau_i$. The number of exponentials used in the fit was chosen so that the $\chi^2$/NDoF value is minimal.
It should be noted that although an approximation of the pulse shape by a sum of exponentials is commonly used \cite{mikhailik07, moszynski05, zdesenko05}, especially at the beginning of the scintillation pulse a non-exponential behavior is expected. In this region the time evolution is influenced by the scintillation quenching mechanism caused by the interaction of self-trapped excitons (STEs) \cite{roth15}.

The values obtained from the fits to the average pulse shapes at different temperatures are shown in Tab.~\ref{tab:decaytimes}. In comparison to the decay times in Ref.~\onlinecite{mikhailik07} we have found an additional long-lasting component in the ms range that starts to become significant for temperatures $\lesssim$40\,K. In general, such a lengthening of the observed decay times could be caused by pileup events. In this case one would expect a difference in pulse shapes from measurements with varying excitation rate. However, we found no difference between the pulse shapes using only the internal $^{241}$Am source and using additional external $^{57}$Co or $^{137}$Cs sources. In addition, a similarly long component was also observed in measurements at mK temperatures using low-temperature light detectors \cite{distefano03}. The scintillation decay times of $\tau_1=0.3\pm0.1$\,ms, $n_1=70\pm15$\% and $\tau_2=2.5\pm1$\,ms found in Ref.~\onlinecite{distefano03} at a temperature of $\sim$20\,mK are in good agreement with the values at 4.2\,K in the present work (see Tab.~\ref{tab:decaytimes}). Furthermore, we note that in the isostructural PbWO$_4$ crystals, a ms decay component was also observed below 50\,K \cite{nagornaya09}. This might point to some common trapping mechanism for both materials at that temperature range. The absence of this long-lasting component in Ref.~\onlinecite{mikhailik07} could be explained by their lower light collection efficiency. Furthermore, their choice of a rather small binning at late times limits the ability to determine slow decay times. Finally, a possible absence of the component in other crystal samples can not be excluded.\\
It is known that there is a noticeable difference between the decay curves of $\alpha$ and $\gamma$ events that allows pulse shape discrimination \cite{mikhailik07,zdesenko05}. In general it was found that $\alpha$ events have faster pulses compared to $\gamma$ events. This difference can be ascribed to the quenching mechanism in CaWO$_4$ \cite{roth15}. Since $\alpha$ particles have a higher ionization density compared to electrons they generate a higher initial density of STEs. This leads to an increased interaction probability for STEs and thus facilitates their non-radiative destruction via the F\"orster interaction. Therefore, at the beginning of the pulse when the STE density is highest, a shortening of the average lifetime of the STE population and hence a shortening of the decay time is expected for $\alpha$ particles with respect to $\gamma$ events.
This effect can be observed in the average pulses shown in Fig.~\ref{fig:pulses_TUM}(a)-(d) and explains the additional short component obtained for $\alpha$ events with respect to $\gamma$'s (see Tab.~\ref{tab:decaytimes}). \\
Comparing the pulse shapes of the two different samples under study we find that at low temperatures the TUM crystal has slightly slower pulses than the sample from GPI. The emission kinetics and especially the long decay component are influenced by trapping centers in the crystal. Therefore, a higher density of trapping centers in the TUM crystal might account for the slower pulses. \\
We note that due to the slow pulses and limited statistics we were not able to determine any possible fast ($\sim$1\,\textmu s) decay component for temperatures $\leq$40\,K as it was observed in Ref.~\onlinecite{mikhailik07}. We also remark that, in addition to the quenching mechanism, some difference between the number of observed components for $\alpha$ and $\gamma$ events may also result from the larger binning which was chosen for the latter due to the lower number of detected photons. This could in particular explain why we observe only two components for $\gamma$ events compared to four components for $\alpha$'s at 4.2\,K.

\begin{table*}[htb]
\begin{tiny}
\caption{\label{tab:decaytimes}
Scintillation components with decay time $\tau$ and the fraction of photons n at various temperatures for $\alpha$ and $\gamma$ events from the $^{241}$Am source. The values were determined by fitting the average pulse shapes with a sum of exponentials (see Eq.~(\ref{eq:pulses})). In comparison we show the values from Ref.~\onlinecite{mikhailik07} measured for $\gamma$ quanta and $\alpha$ particles using a $^{60}$Co and a $^{241}$Am source, respectively.\footnote{Fits in Ref.~\onlinecite{mikhailik07} were carried out with a function of the form $\sum_{i}{A_i\exp(-t/\tau_i)}$; here we have converted the amplitudes $A_i$ to $n_i$ as per our Eq.~(\ref{eq:pulses}).}
}
\begin{ruledtabular}
\begin{tabular}{cccccccc}
 & & \multicolumn{2}{c}{\textbf{TUM crystal}} & \multicolumn{2}{c}{\textbf{GPI crystal}} & \multicolumn{2}{c}{\textbf{Ref.~\onlinecite{mikhailik07}}} \\
\hline
& & $\tau$ (\textmu s) & $n$ (\%) & $\tau$ (\textmu s) & $n$ (\%) & $\tau$ (\textmu s) & $n$ (\%) \\
\hline
\multirow{5}{*}{295\,K} & \multirow{3}{*}{$\alpha$ particles} & 0.33$\pm$0.01 & 3.6$\pm$0.1 & 0.46$\pm$0.01 & 5$\pm$0.1 & - & - \\
& &  2.36$\pm$0.03 & 18.8$\pm$0.2 &  2.79$\pm$0.04 & 21.6$\pm$0.2  & 1.0$\pm$0.2 & 7.2 \\ 
& &  8.72$\pm$0.01 & 77.6$\pm$0.2  &  8.64$\pm$0.02 & 73.4$\pm$0.3  & 8.6$\pm$0.3 & 92.8 \\ 
\cline{2-8}
& \multirow{2}{*}{$\gamma$ quanta} &  1.48$\pm$0.1 & 5.1$\pm$0.3 &  1.1$\pm$0.06 & 3.9$\pm$0.2  & 1.4$\pm$0.2 & 6.1 \\ 
& & 8.48$\pm$0.03 & 94.9$\pm$0.4 & 8.09$\pm$0.02 & 96.1$\pm$0.2  & 9.2$\pm$0.3 & 93.9 \\ 
\hline
\multirow{7}{*}{77\,K} & \multirow{4}{*}{$\alpha$ particles} & 0.42$\pm$0.01 & 3$\pm$0.1 &  0.59$\pm$0.01 & 4.2$\pm$0.1 & - & - \\
& &  4.11$\pm$0.05 & 16.2$\pm$0.1 & 4.34$\pm$0.04 & 22.4$\pm$0.1 & 3.2$\pm$0.3 & 16.2\\ 
& & 16.73$\pm$0.02 & 80.8$\pm$0.2 &  15.7$\pm$0.03 & 73.3$\pm$0.2  & 16.6$\pm$0.4 & 83.8\\ 
\cline{2-8}
& \multirow{3}{*}{$\gamma$ quanta} &  3.30$\pm$0.13 & 7.2$\pm$0.3 & 3.15$\pm$0.1 & 10.4$\pm$0.4 & 2.1$\pm$0.3 & 15.2 \\ 
& &  16.51$\pm$0.04 & 92.8$\pm$0.3 &  15.31$\pm$0.05 & 89.6$\pm$0.4  & 17.6$\pm$0.3 & 84.8\\ 
\hline
\multirow{7}{*}{4.2\,K} & \multirow{4}{*}{$\alpha$ particles} & 21.76$\pm$0.8 & 2.5$\pm$0.1 & 18.01$\pm$0.55 & 2.6$\pm$0.1  & 1.0$\pm$0.2 & 12.9\\ 
& &  151.7$\pm$10.3 & 10.5$\pm$1.4 &  124.6$\pm$3.97 & 13.1$\pm$0.6  & - & - \\
& &  371.1$\pm$6.2 & 68.2$\pm$1.1  & 352.28$\pm$2.91 & 71.5$\pm$0.5 & 330$\pm$40 & 87.1\\ 
& & 2057$\pm$92 & 18.8$\pm$0.3 & 2007.48$\pm$72.43 & 12.8$\pm$0.2 & - & - \\
\cline{2-8}
& \multirow{3}{*}{$\gamma$ quanta} & - & -  & - & - & 0.9$\pm$0.2 & $>$27.1\\ 
& &  369.2$\pm$4.6 & 56.1$\pm$0.8 &  347.07$\pm$3.02 & 68.2$\pm$0.7  & 480$\pm$60 & 72.9\\ 
& & 2955$\pm$222 & 43.9$\pm$0.8 & 2359.63$\pm$138.54 & 31.8$\pm$0.3 & - & - \\
\end{tabular}
\end{ruledtabular}
\end{tiny}
\end{table*}

\begin{figure}
\subfigure[]{\includegraphics[width=8.5cm]{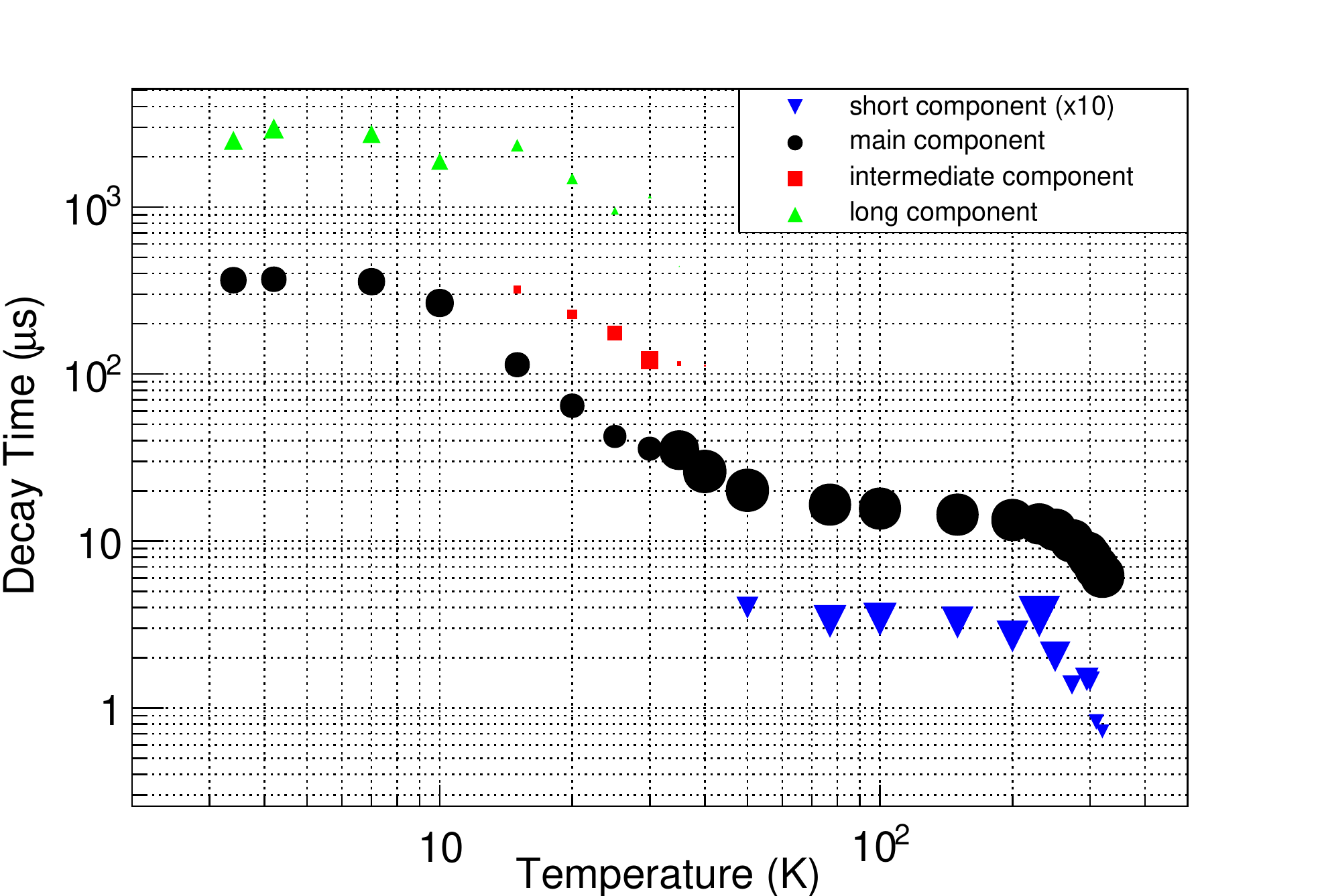}}
\subfigure[]{\includegraphics[width=8.5cm]{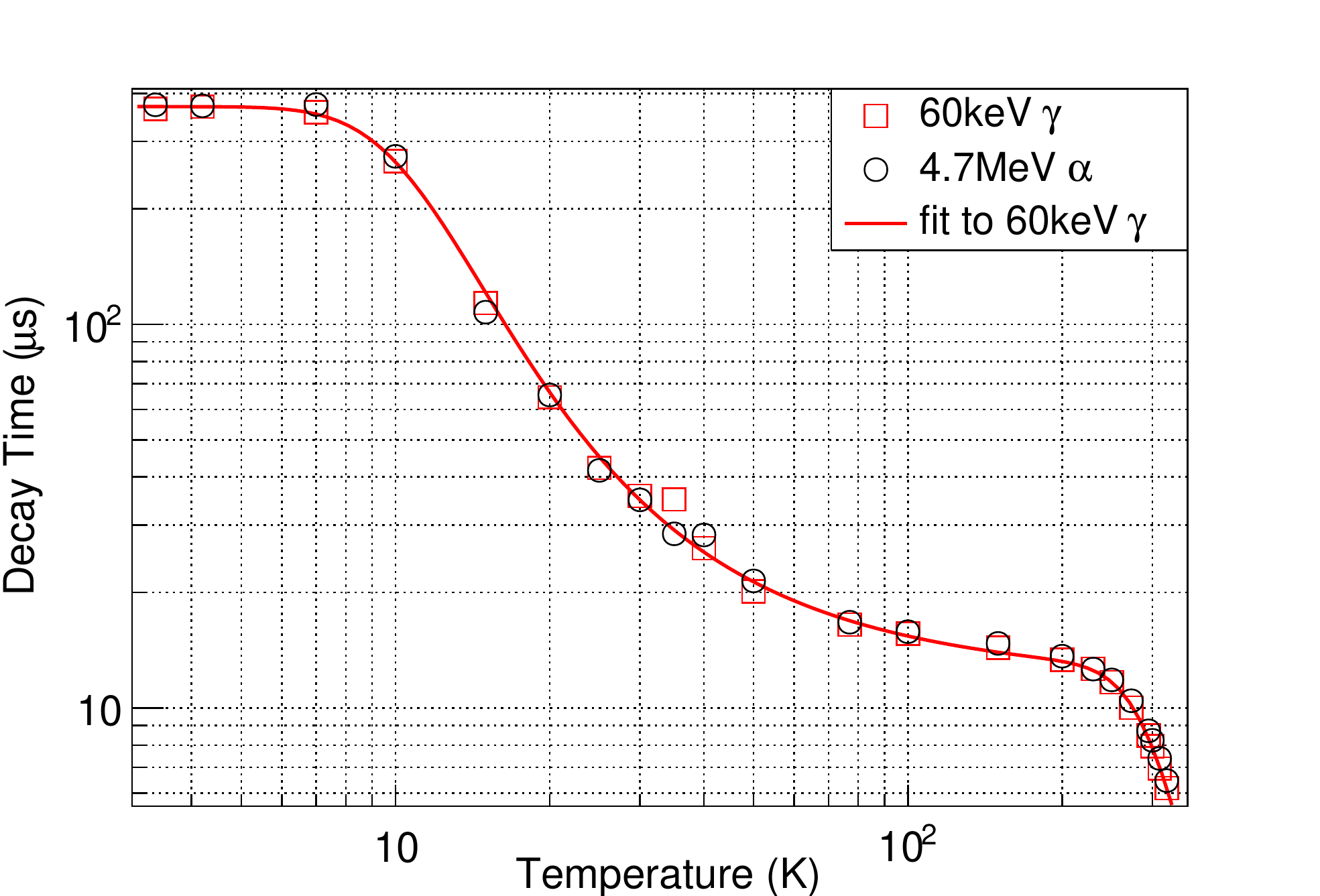}}
\caption{\label{fig:decaytime_TUM}
(a): Decay times versus temperature measured with the TUM crystal under excitation with 60\,keV $\gamma$-rays. The size of the markers corresponds to the fraction of photons that are contained within each component. The symbols of the short component were enlarged 10$\times$ for better visibility.\\
(b): Main decay times versus temperature for excitation with 60\,keV $\gamma$-rays and 4.7\,MeV $\alpha$ particles measured with the TUM crystal. The line shows the best fit to the data of the $\gamma$ quanta according to Eq.~(\ref{eq:decaytime}). Error bars are smaller than symbols.}
\end{figure}

Figure \ref{fig:decaytime_TUM}(a) shows the temperature dependence of the decay times obtained for $\gamma$ events measured with the TUM crystal. The fraction of photons that is contained in each component is proportional to the area of the symbols in the plot, where those of the short component were enlarged by a factor of 10 for better visibility. From the plot one can identify the main component of the pulse shape which comprises the largest fraction of the light yield at all temperatures. This component is usually ascribed to the intrinsic scintillation caused by recombination of STEs (see Ref.~\onlinecite{nikl08} and references therein). We further note that an intermediate component appears between 15-40\,K which was also found in Refs.~\onlinecite{mikhailik07, kraus05} under excitation with $\alpha$ particles. As the pulse shape is only approximated by a sum of exponentials this component does not necessarily correspond to a physical process. In Refs.~\onlinecite{mikhailik07, kraus05} it was assigned to the luminescence from delayed recombination processes. 

It is established that the temperature dependence of the main decay time in CaWO$_4$ can be described with a three-level model \cite{mikhailik07}. In addition, the model has previously been used to describe the decay times and emission intensities as a function of temperature in PbWO$_4$ \cite{nikl97}. The increase of the decay time is ascribed to a metastable level which can be thermally depopulated at high temperatures but slows down the emission process at lower temperatures. The key parameters of this model are the probabilities for radiative decay from the two excited levels denoted by $k_1$ and $k_2$, respectively, and the energy splitting $D$ of the two levels. Furthermore, the model includes a term for non-radiative quenching to the groundstate (STE-recombination via phonon production) described by the decay rate $K$ and the energy barrier $\Delta E$. The decay time $\tau$ as a function of temperature $T$ can then be calculated by the following equation:
\begin{equation}
1/\tau=\frac{k_1+k_2\exp(-D/k_BT)}{1+\exp(-D/k_BT)}+K\exp(-\Delta E/k_BT)
\label{eq:decaytime}
\end{equation}
where $k_B$ denotes the Boltzmann constant.\\
We have used Eq.~(\ref{eq:decaytime}) to fit the temperature evolution of the main decay time that was derived for $\gamma$ events in the TUM crystal (see Fig.~\ref{fig:decaytime_TUM}(b)). The parameters obtained from this fit are summarized in Tab.~\ref{tab:fit_parameters}.
The value of the energy splitting $D$, and the emission probabilities $k_1$ and $k_2$ are comparable to the values quoted in Ref.~\onlinecite{mikhailik07}. However, for the energy barrier for non-radiative quenching $\Delta E$, and the associated decay rate $K$ we find lower values than those given in Ref.~\onlinecite{mikhailik07}. We note that we observe no differences between the main decay time of $\alpha$ and $\gamma$ events and hence no influence of the ionization density. This suggests that the component arrives solely from the recombination of STEs and is not influenced by the quenching mechanism from the interaction of STEs.

\begin{table}[htb]
\caption{\label{tab:fit_parameters} Parameters obtained by fitting Eq.~(\ref{eq:decaytime}) to the decay time as a function of temperature as shown in Fig.~\ref{fig:decaytime_TUM}(b). For comparison the values obtained in Ref.~\onlinecite{mikhailik07} are also given.}
\begin{ruledtabular}
\begin{tabular}{llll}
 & Present work & Ref.~\onlinecite{mikhailik07} \\
\colrule
 k$_1$ (s$^{-1}$) & (2.71$\pm$0.02)$\times$10$^{3}$ & 3.0$\times$10$^3$ \\
 k$_2$ (s$^{-1}$) & (1.679$\pm$0.003)$\times$10$^{5}$ & 1.1$\times$10$^5$ \\
 K (s$^{-1}$) & (2.7$\pm$0.2)$\times$10$^{8}$ & 8.6$\times$10$^9$ \\
 D (meV) & 4.34$\pm$0.01 & 4.4 \\
 $\Delta$E (meV) & 223$\pm$2 & 320  \\
\end{tabular}
\end{ruledtabular}
\end{table}%

\subsection{Relative Light Yield}
\begin{figure}
\includegraphics[width=8.5cm]{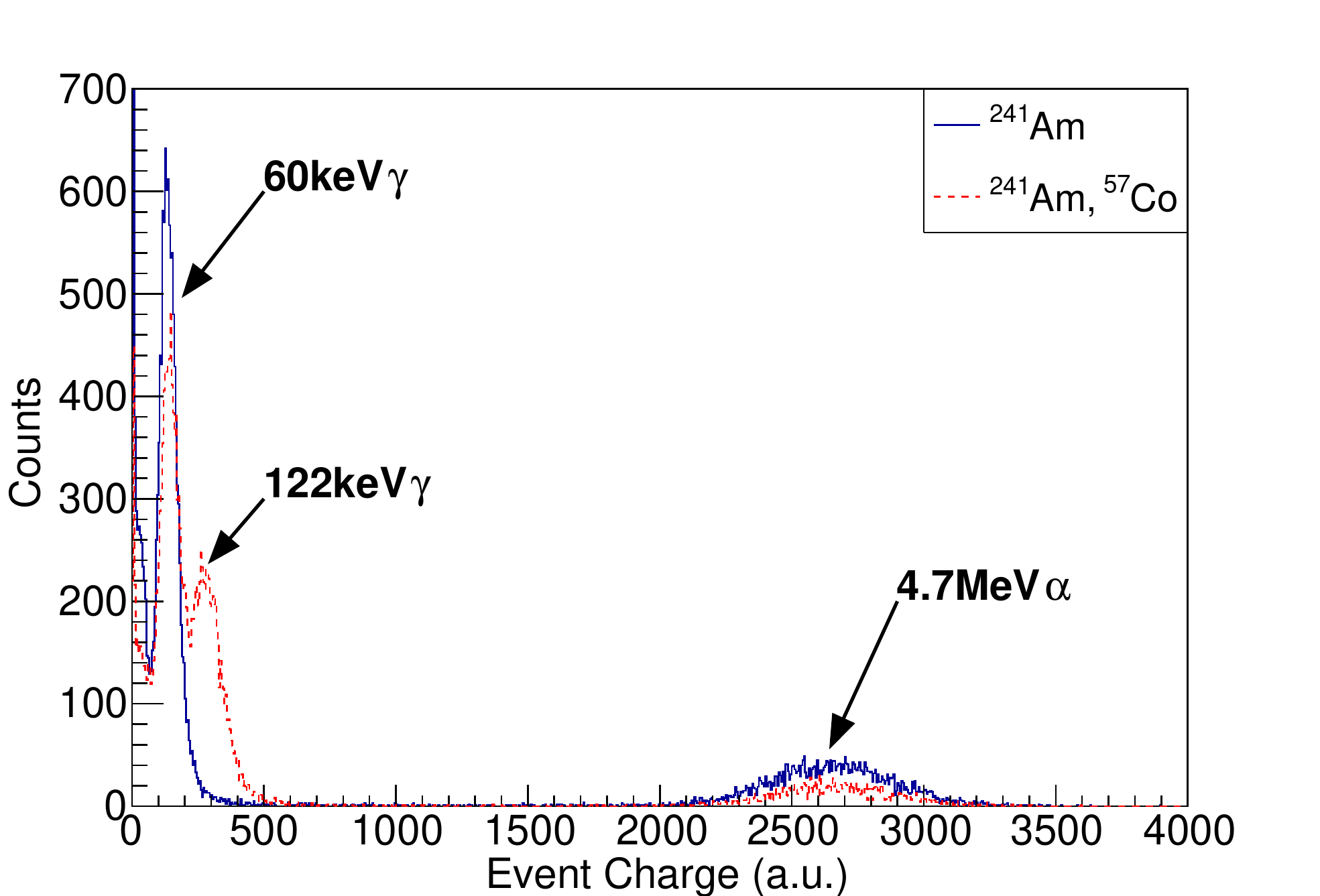}
\caption{\label{fig:spectrum_TUM} Spectra measured with the TUM crystal at 77\,K using the built-in $^{241}$Am source and an external $^{57}$Co source. The $^{241}$Am spectrum shows a narrow peak from the 60\,keV $\gamma$-rays and a broad peak due to degraded $\alpha$ particles with a mean energy of 4.7\,MeV. With the $^{57}$Co source an additional peak from the 122\,keV $\gamma$-rays is visible.}
\end{figure}

As an example, Fig.~\ref{fig:spectrum_TUM} depicts two spectra that were recorded with the TUM crystal at 77\,K. One spectrum was obtained only with the internal $^{241}$Am source while the other one was recorded using the additional $^{57}$Co source. The  $^{241}$Am spectrum shows the photopeak from the 60\,keV $\gamma$-rays and a broader peak from $\alpha$ particles. A measurement of the $\alpha$ spectrum from the source with a silicon detector showed that the $\alpha$ peak appears at a lower energy of 4.7\,MeV compared to the expected energy of 5.5\,MeV and is broadened with a FWHM of 0.5\,MeV. These degradation effects result from a thin window protecting the source. In the spectrum in Fig.~\ref{fig:spectrum_TUM} with the additional $^{57}$Co source another peak is visible which mainly results from the 122\,keV $\gamma$-rays. The 136\,keV $\gamma$ quanta that are also emitted with a factor of $\sim$8 smaller branching ratio cannot be separated due to the limited resolution. For each temperature $T$ we calculated the relative light yield $RLY$ which represents the number of photons produced per unit energy $E$ by a given particle. The value is normalized to that of 60\,keV $\gamma$-rays at 295\,K and was calculated by the following equation: 
\begin{equation}
RLY(T)=\frac{\mu(T)/E}{\mu_{60\text{keV}}(295\,\text{K})/60\,\text{keV}}
\label{eq:LY}
\end{equation}
Here $\mu$ is the position of the corresponding peak in the spectrum (see Fig.~\ref{fig:spectrum_TUM}) derived from a Gaussian fit. The value $\mu_{60\text{keV}}$(295\,K) denotes the position of the 60\,keV peak at 295\,K. 

\begin{figure}
\includegraphics[width=8.5cm]{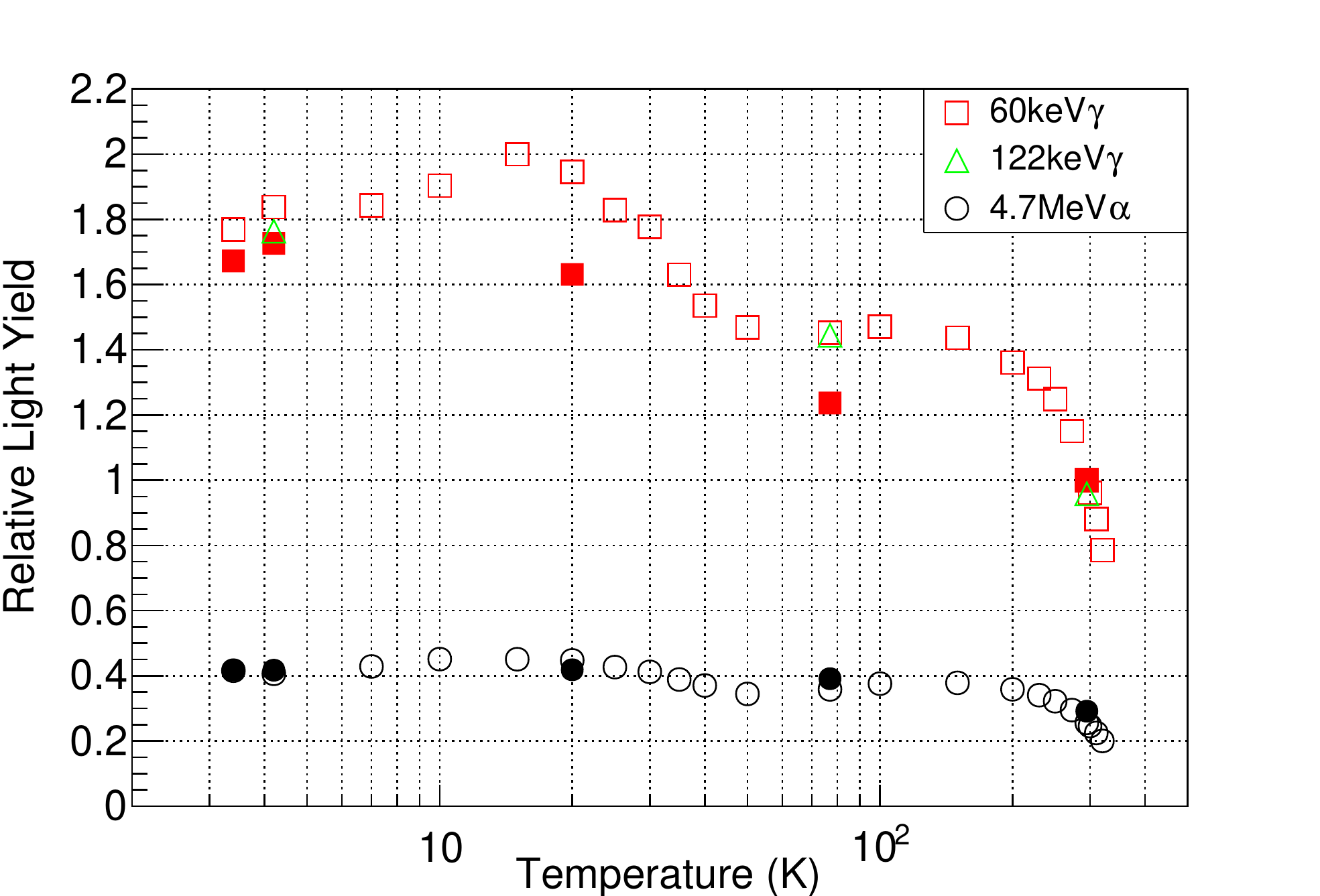}
\caption{\label{fig:LY} Relative light yield ($RLY$) of different particles as a function of temperature. Open symbols show the results for the TUM crystal and filled symbols for the sample from GPI. The measurements were carried out with an $^{241}$Am source emitting 60\,keV $\gamma$-rays and degraded $\alpha$ particles with an energy of 4.7\,MeV. In some measurements, an additional $^{57}$Co $\gamma$-ray source (122\,keV) was used. Error bars are smaller than symbols.}
\end{figure}

For the TUM crystal we recorded a full sweep at 22 different temperatures while for the crystal provided by GPI we only recorded spectra at 5 different temperatures. Because of the unexpected slow ($\sim$ms) scintillation component that appears at temperatures $\lesssim$40\,K (see section \ref{sec:pulse_shapes}) some photons are not detected within the acquisition window of 3\,ms. Therefore, at temperatures $\leq$20\,K the light yield had to be corrected. First the average pulse shape was fitted with a sum of exponentials (see Eq.~(\ref{eq:pulses})). The fit was then used to calculate a correction factor $\epsilon$ for the light yield using the following equation:
\begin{equation}
\epsilon=\frac{\int_0^\infty \! f(t) \, \mathrm{d}t}{\int_0^{t_{\text{acq}}} \! f(t) \, \mathrm{d}t}
\end{equation}
Here $f(t)$ is the fitted pulse and $t_{\text{acq}}$ is the length of the acquisition window. Depending on the crystal sample and temperature the calculated correction factor is between 2-6\% for $\alpha$ particles and between 4-17\% for $\gamma$ events.\\
The results for the relative light yield $RLY$ as a function of temperature for both crystals, after applying the correction factor $\epsilon$, are shown in Fig.~\ref{fig:LY}.
The curves can be qualitatively understood by the semi-empirical model in Ref.~\onlinecite{mikhailik10} and a recently developed comprehensive microscopic model of the scintillation mechanism in CaWO$_4$\cite{roth15,roth12}. Below 320\,K, $RLY$ increases steeply due to the reduced probability of thermal quenching. At temperatures $\sim$200-40\,K $RLY$ stays approximately constant. For lower temperatures the mobility of excitons starts to decrease which reduces their interaction probability and hence increases $RLY$. Below $\sim$5\,K excitons are essentially immobile and $RLY$ stays approximately constant. Similarily to Ref.~\onlinecite{mikhailik07} we observe a maximum of $RLY$ around $\sim$15\,K. A detailed explanation of this feature is given in Ref.~\onlinecite{roth12}. 
The total increase of the light yield for $\alpha$ particles at 9\,K compared to 295\,K quoted in Ref.~\onlinecite{mikhailik10} is 1.82. This is compatible with the value of 1.77$\pm0.04$ we observed at 10\,K for the TUM crystal.\\ 
For the sample from GPI we found that the total increase of $RLY$ under both $\alpha$ and $\gamma$ excitation is $\sim$5-15\% lower than for the TUM crystal. This difference between the two samples is most pronounced when the temperature decreases from 295\,K to 77\,K (see Fig.~\ref{fig:LY}). As explained above the temperature dependence of $RLY$ in this range is dominated by thermal quenching. A larger probability for non-radiative recombination at room temperature will lead to a larger increase of $RLY$ at lower temperatures when the process becomes negligible. The probability for thermal quenching is $\sim\exp(-\Delta E/k_BT)$ (see Eq.~(\ref{eq:decaytime})), hence a lower value of $\Delta$E can explain the larger increase of $RLY$ in the TUM crystal. \\
The similar value of $RLY$ for 60\,keV $\gamma$-rays from the $^{241}$Am source and that for 122\,keV $\gamma$ quanta from the external $^{57}$Co source is in agreement with a linear behavior of the scintillator. The known non-linearity of CaWO$_4$ scintillators which is $\lesssim$5\% in this energy range \cite{moszynski05,lang09} is below the accuracy of our measurement.

\subsection{$\alpha$/$\gamma$ Light-Yield Ratio}
\begin{figure}
\includegraphics[width=8.5cm]{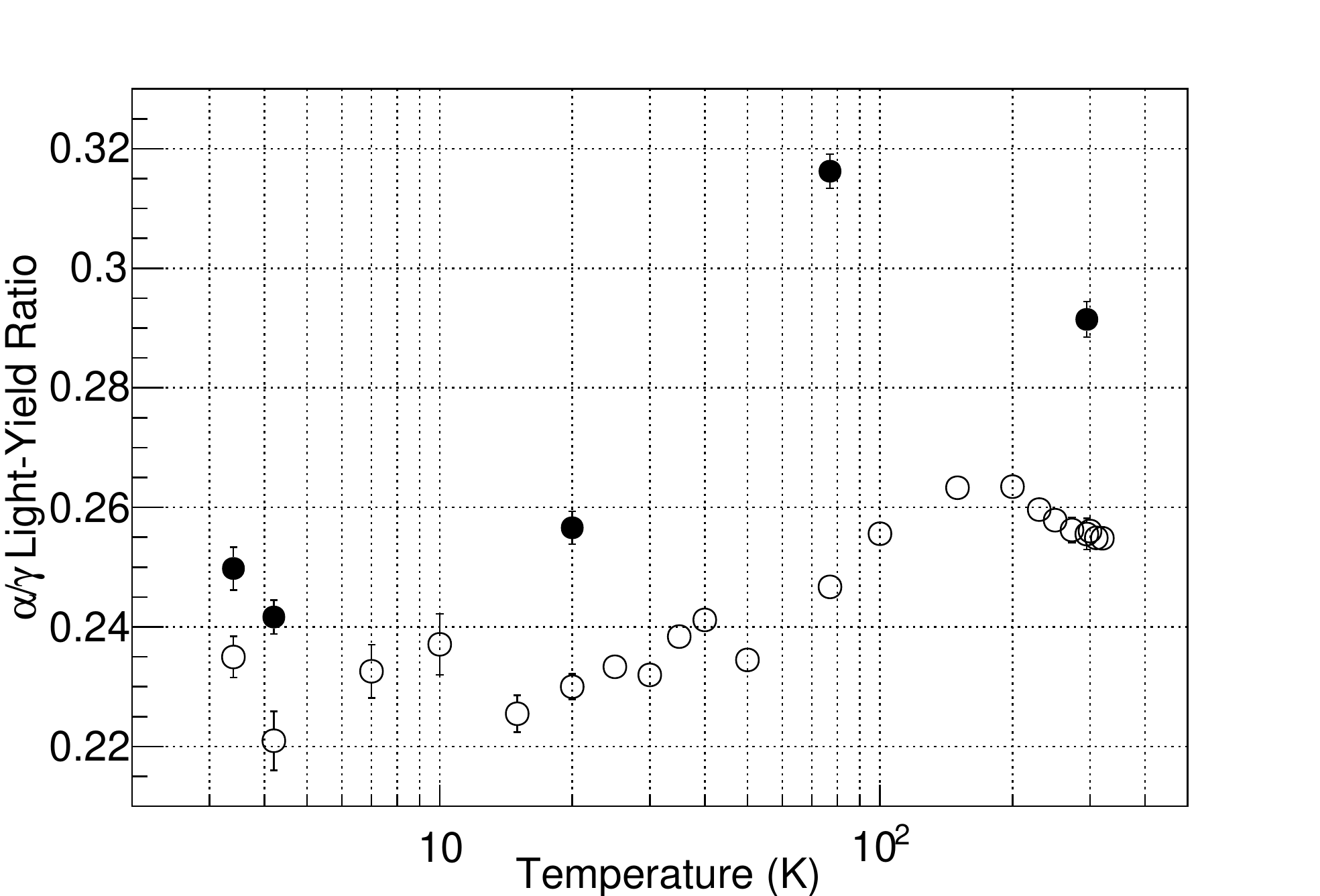}
\caption{\label{fig:QF} Temperature dependence of the $\alpha$/$\gamma$ light-yield ratio ($Q$) determined with the $^{241}$Am source. Open symbols refer to the measurement with the TUM crystal and filled symbols to the GPI crystal.}
\end{figure}

An important characteristic of a scintillator is the $\alpha$/$\gamma$ light-yield ratio $Q$ which quantifies the different light yield of $\alpha$ and $\gamma$ events. This quantity can be used for particle discrimination in scintillating bolometers and therefore to suppress background from $\alpha$ events in experiments searching for dark matter or the neutrinoless double beta decay \cite{alessandrello98,bobin97}. We define $Q$ as the ratio between the light yield of a $^{241}$Am $\alpha$ event and that of a 60\,keV $\gamma$ event at temperature $T$ as calculated by the following equation: 
\begin{equation}
Q(T)=\frac{\mu_{\alpha}(T)/4.7\,\text{MeV}}{\mu_{60\text{keV}}(T)/60\,\text{keV}}
\end{equation}
Here $\mu_\alpha$ and $\mu_{60\text{keV}}$ are the positions of the peaks in the spectrum coming from the $^{241}$Am $\alpha$ particles and $\gamma$ quanta, respectively.  
A closely related quantity is the $\alpha$/$\gamma$ quenching factor which is, however, defined for equal energy deposits. Both quantities would be the same if the scintillator behaved linearly.
Fig.~\ref{fig:QF} shows the value of $Q$ as a function of temperature. Depicted are the data points after correction of the light yield for the limited acquisition window.
For the TUM crystal at 295\,K we have obtained $Q=0.256\pm0.003$. This is compatible with the value of $Q=0.23\pm0.02$ which we have calculated for $\alpha$ particles of this energy from the following equation quoted in Ref.~\onlinecite{zdesenko05}:
\begin{equation}
Q=a+b\cdot E,
\label{eq:QF}
\end{equation}
with $a=0.129\pm0.012$, $b=0.021\pm0.003$ and the energy $E$ in MeV.
Dependent on temperature the value of $Q$ of the GPI crystal is $\sim$6-14\% higher than that of the TUM crystal. This reduced strength of the quenching effect in the GPI crystal can be explained by a smaller density of STEs\cite{roth15} caused by a larger defect density. It should be noted that this is not necessarily in conflict with a smaller concentration of another type of defect center that could account for the shorter scintillation decay time. A better understanding of the different types of defect centers present in the crystals would require further investigation using e.g. spectrally resolved luminescence or thermoluminescence measurements.
For both samples we observe that the temperature dependence of $Q$ shows a non-monotonic behavior. As temperature decreases, $Q$ first increases and then decreases. For the TUM crystal the maximum value of $Q$ is between 150-200\,K while it stays approximately constant below 30\,K. For both crystals the value at the lowest temperature is $\sim$8-15\% lower than the room temperature value. Such a temperature dependence is predicted when the quenching is described by non-radiative destruction of STEs via the F\"orster interaction \cite{roth15}. The lower value of $Q$ at low temperatures can be interpreted as an increase of this interaction length described by the F\"orster radius.

\section{Summary and Conclusion}
We have studied the scintillation properties of CaWO$_4$ crystals in the temperature range 320-3.4\,K. For the first time we have measured the temperature dependence of the light yield of $\gamma$ quanta and found that it differs from those of $\alpha$ particles. This translates into a temperature dependence of the $\alpha$/$\gamma$-ratio of the light yield which is $\sim$8-15\% smaller at low temperatures compared to room temperature. This ratio is important to discriminate background from $\alpha$ events in experiments searching for dark matter or neutrinoless double beta decay with scintillating bolometers. Our measurements show that the $\alpha$/$\gamma$ light-yield ratio can not be inferred from measurements at room temperature unless the temperature dependence can be correctly modeled. \\ 
Investigating the scintillation decay times, we have observed a long component in the ms range at temperatures $\lesssim$40\,K that is contributing significantly to the total light yield. In this regard our systematic study confirms measurements carried out with scintillating bolometers at lower temperatures. For rare-event searches with bolometers based on CaWO$_4$, long scintillation decay times are not problematic because of the expected low event rates. In addition, the intrinsic decay time of low-temperature light detectors is also in the ms range \cite{distefano03}.\\ 
The comparison of two different samples showed differences in the temperature evolution of the light yield and the decay times which we ascribe to different concentrations of crystal defects that act as trapping centers or cause reduction of the light yield.
The observed increase of the light yield with decreasing temperature is especially relevant for dark matter searches where very low recoil energies $\mathcal{O}$(10\,keV) are expected. To improve the production of crystals for these experiments it is therefore important to better understand which crystal properties influence the light yield at low temperatures. 

\begin{acknowledgments}
This work has been funded in Canada by NSERC (Grant SAPIN 386432), CFI-LOF and ORF-SIF (project 24536).
The research was also supported by the DFG cluster of excellence 'Origin and Structure of the Universe', the Helmholtz Alliance for Astroparticle Physics, and the Maier-Leibnitz-Laboratorium (Garching).
We thank Federica Petricca and Franz Pr\"obst of MPP Munich for providing the GPI crystal sample.

\end{acknowledgments}


\providecommand{\noopsort}[1]{}\providecommand{\singleletter}[1]{#1}%

\end{document}